\documentclass[aps,prd,12pt,showpacs,notitlepage,nofootinbib,tightrnlines]{revtex4-1}
\usepackage{amsmath}
\usepackage{bm}
\usepackage{times}
\usepackage{braket}
\usepackage{color,graphicx}
\usepackage{slashed}
\usepackage{hyperref}
\newcommand{\beq}{\begin{eqnarray}}
\newcommand{\eeq}{\end{eqnarray}}
\newcommand{\non}{\nonumber\\ }

\def\lsim{ {\ \lower-1.2pt\vbox{\hbox{\rlap{$<$}\lower6pt\vbox{\hbox{$\sim$}}}}\ } }
\def\gsim{ {\ \lower-1.2pt\vbox{\hbox{\rlap{$>$}\lower6pt\vbox{\hbox{$\sim$}}}}\ } }

\definecolor{Red}{rgb}{1.,0.,0.}

\definecolor{Blue}{rgb}{0.,0.,1.}

\definecolor{nicered}{rgb}{0.7,0.1,0.1}
\definecolor{nicegreen}{rgb}{0.1,0.5,0.1}

\hypersetup{colorlinks,citecolor=nicegreen,linkcolor=nicered}

\begin{document}

\title{ Axial-vector $f_1(1285)-f_1(1420)$ mixing and $B_s \to J/\psi
(f_1(1285),  f_1(1420))$ decays}
\author{Xin~Liu}
\email[Electronic address:]{liuxin.physics@gmail.com}
\affiliation{School of Physics and Electronic Engineering,\\
Jiangsu Normal University, Xuzhou, Jiangsu 221116, People's Republic of China}

\author{Zhen-Jun~Xiao}
\email[Electronic address:]{xiaozhenjun@njnu.edu.cn}
\affiliation{Department of Physics and Institute of Theoretical Physics,\\
Nanjing Normal University, Nanjing, Jiangsu 210023, People's Republic of China}

\date{\today}

\begin{abstract}
Inspired by the very recent LHCb measurements of $B_s \to J/\psi f_1(1285)$
and the good agreement between the perturbative QCD predictions and the data for many $B \to J/\psi V$ decays,
we here investigate the $B_s \to J/\psi f_1(1285)$ and $B_s \to J/\psi f_1(1420)$
decays for the first time by employing the perturbative QCD (pQCD) 
approach,
in which the $1^3P_1$ states $f_1(1285)$ and $f_1(1420)$ are believed to be the mixture of
flavor singlet $f_1$ and octet $f_8$ or of quark-flavor states $f_{1q}$ and $f_{1s}$.
We show that the pQCD predictions for the branching ratio of $B_s \to J/\psi f_1(1285)$
agree well with the data within errors for the mixing angle $\theta_{^3P_1} \approx 20^\circ
(\phi_{^3P_1} \approx 15^\circ)$
between $f_1 (f_{1q})$ and $f_8 (f_{1s})$ states. Furthermore, the branching ratio of $B_s \to J/\psi f_1(1420)$ and the large
transverse polarization fractions in these two considered channels
are also predicted and will be tested by the LHC and the forthcoming Super-B factory experiments.
Based on the decay rates of $B_s \to J/\psi f_1(1285)$ and $B_s \to J/\psi f_1(1420)$ decay modes
predicted in the pQCD approach, the extracted mixing angle between $f_1(1285)$ and $f_1(1420)$
is basically consistent with currently available experimental measurements and lattice
QCD analysis within still large theoretical errors.

\end{abstract}

\maketitle

%
%

Very recently, the LHCb Collaboration located at CERN reported the
first observation of $B_s \to J/\psi f_1(1285)$ decay with
the branching ratio~\cite{Aaij:2013rja},
 \beq
 Br(B_s \to J/\psi f_1(1285))_{\rm Exp.} &=&
 \left ( 7.14 \pm 0.99^{+0.83}_{-0.91} \pm 0.41\right ) \times 10^{-5} \;.
 \label{eq:psif12-ex}
 \eeq
Of course, the accuracy of the above data is expected to be improved
rapidly with the future LHCb and Super-B experiments. By combining
the first measurement of $B_d \to J/\psi f_1(1285)$ channel with this new one,
the mixing angle between the strange and non-strange component of
the wave function of $f_1(1285)$ in the $q\bar q$ structure model is
determined to be $\pm(24.0^{+3.1+0.6}_{-2.6-0.8})^\circ$~\cite{Aaij:2013rja}
for the first time {\it in B meson decays}.

In the quark model, as is well-known, $f_1(1285)$ is treated as a $p$-wave
axial-vector meson with $J^{PC} = 1^{++}$, which is believed to mix with its partner
$f_1(1420)$~\cite{Close:1997nm,Li:2000dy} just like the "$\eta-\eta'$" mixing
in the pseudoscalar sector.
Up to now, many discussions have been presented on the
mixing angle $\theta_{^3P_1}$ or $\phi_{^3P_1}$ of $f_1(1285)-f_1(1420)$ mixing,
in the framework of the two popular mixing schemes: i.e.,
the so-called singlet-octet(SO) basis and the quark-flavor(QF)
basis~\cite{Gidal:1987bn,Li:2005eq,
Carvalho:2002fh,Yang:2007zt,Cheng:2007mx,Yang:2008xw,Cheng:2008gxa,
Yang:2010ah,Cheng:2011pb,Dudek:2011tt,Stone:2013eaa,Dudek:2013yja,Cheng:2013cwa}.
One of the most important reasons is that the mixing
angle $\theta_{^3P_1}$ or $\phi_{^3P_1}$ can be utilized to constrain
the magnitude of the mixing angle $\theta_{K_1}$ of axial-vector
$K_{1}(1270)-K_{1}(1400)$ system~\cite{Cheng:2011pb}, which is a very
special mixing between two distinct types of axial-vector mesons
$K_{1A}$($1^3P_1$) and $K_{1B}$($1^1P_1$).

In the SO basis, the axial-vector $f_1(1285)-f_1(1420)$ mixing can be written
in the form of ~\cite{Beringer:1900zz},
\beq
\left(
\begin{array}{c} f_1(1285)\\ f_1(1420) \\ \end{array} \right ) &=&
  \left( \begin{array}{cc}
 \hspace{0.28cm}\cos{\theta_{^3P_1}} & \sin{\theta_{^3P_1}} \\
-\sin{\theta_{^3P_1}} & \cos\theta_{^3P_1} \end{array} \right )
 \left( \begin{array}{c}  f_1\\ f_8 \\ \end{array} \right )\;,
 \label{eq:f1-f8-mix}
 \eeq
with the SO states $f_1= (u \bar u + d \bar d + s \bar s)
/\sqrt{3}$ and $f_8= (u \bar u + d \bar d - 2 s \bar s)/ \sqrt{6}$.
While, in the QF basis, the $f_1(1285)-f_1(1420)$ mixing
can be written as the following pattern~\cite{Beringer:1900zz},
\begin{eqnarray}
\left\{ \begin{array}{ll}
f_1(1285) =  \cos\phi_{^3P_1} f_{1q} + \sin\phi_{^3P_1} f_{1s}\; &  \\
f_1(1420) =  \sin\phi_{^3P_1} f_{1q} - \cos\phi_{^3P_1} f_{1s}\; &   \\ \end{array} \right.
\label{eq:fq-fs-mix}
\end{eqnarray}
with the QF states $f_{1q}=(u\bar u + d \bar d)/\sqrt{2}$
and $f_{1s} = s \bar s$. The QF mixing angle $\phi_{^3P_1}$ is
related to the SO mixing angle $\theta_{^3P_1}$ by the
relation $\phi_{^3P_1} = \theta_i - \theta_{^3P_1}$, where $\theta_i$
is the "ideal" mixing angle with $\theta_i = 35.3^\circ$. Therefore,
$\phi_{^3P_1}$ measures the deviation from ideal mixing.

Though the $f_1(1285)$ mixing angle has been preliminarily determined through the
$B_{d/s} \to J/\psi f_1(1285)$ decays in the QF basis by the LHCb Collaboration,
it is necessary to point out that the assumption of exact SU(3) flavor symmetry
on the decay amplitudes of $B_{d/s} \to J/\psi f_1(1285)$ has been adopted
there~\cite{Aaij:2013rja}. In fact, at the
theoretical aspect, the contributing components in the above mentioned
$B_d \to J/\psi f_1(1285)$ and $B_s \to J/\psi f_1(1285)$ decays at the quark
level should be the QF states $f_{1q}$ and $f_{1s}$ respectively,
whose behavior may be rather different
because of the breaking of SU(3) flavor symmetry for $f_{1q}$ and $f_{1s}$.
Consequently, the resultant mixing angles may considerably shift away
from the expected values.

\begin{figure}[!!t]
\centering
\begin{tabular}{l}
\includegraphics[width=0.8\textwidth]{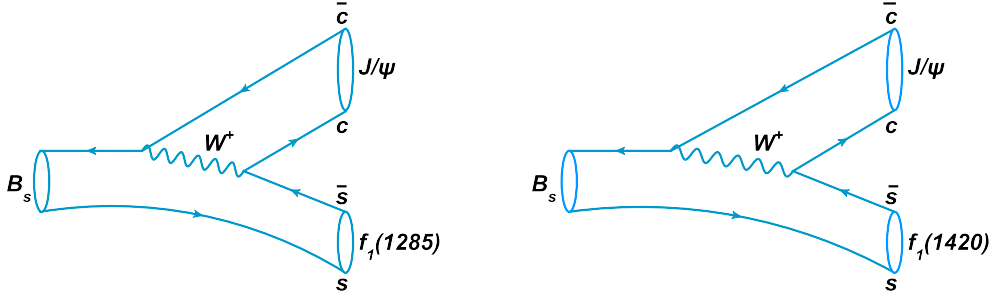}
\end{tabular}
\caption{(Color online) Leading quark-level Feynman diagrams
for the $B_s \to J/\psi f_1(1285)$(left)
and $B_s \to J/\psi f_1(1420)$(right) decays.}
  \label{fig:fig1}
\end{figure}

It may be very interesting to study the mixing angle of $f_1(1285)-f_1(1420)$
mixing through the same components at the quark level, for example,
the $f_1(1285)$ and $f_1(1420)$ mesons are produced through their
strange components in the $B_s$ meson decays, as illustrated in Fig.~\ref{fig:fig1}.
For the $B_s \to J/\psi f_1(1285)$ ( $B_s \to J/\psi f_1(1420)$ ) decay,
the coefficient for the
$s\bar s$ component is $\sin\phi_{^3P_1}$ ( $-\cos\phi_{^3P_1}$ )
in the QF basis, and $\frac{\cos\theta_{^3P_1}}{\sqrt{3}}
- \frac{\sqrt{2} \cdot \sin\theta_{^3P_1}}{\sqrt{3}}$
( $-\frac{\sin\theta_{^3P_1}}{\sqrt{3}}
-\frac{\sqrt{2} \cdot \cos\theta_{^3P_1}}{\sqrt{3}}$ ) in the SO basis, respectively.

One can see that the angle $\phi_{^3P_1}$ of $f_1(1285)-f_1(1420)$ mixing in
the QF basis could be extracted more directly and cleanly
through the following ratio,
 \beq
{\rm R_s^{QF}} &\equiv&   \frac{Br(B_s \to J/\psi f_1(1285))}
  {Br(B_s \to J/\psi f_1(1420))}\non
  &=& \frac{\Phi_{f_1(1285)}\cdot |\sin\phi_{^3P_1}\cdot {\cal A}(B_s \to J/\psi f_{1s})|^2}
  {\Phi_{f_1(1420)}\cdot |-\cos\phi_{^3P_1}\cdot {\cal A}(B_s \to J/\psi f_{1s})|^2}
  = \frac{\Phi_{f_1(1285)}}{\Phi_{f_1(1420)}} \cdot \tan^2\phi_{^3P_1}\;,
  \label{eq:r-s-QF}
  \eeq
where $\Phi_{f_1(1285)}$ and $\Phi_{f_1(1420)}$ are the phase space factors
for $B_s \to J/\psi f_1(1285)$ and $B_s \to J/\psi f_1(1420)$ decays, respectively.
Once the precise measurements for the decay rates of these two channels are available,
one could extract the mixing angle $\phi_{^3P_1}$ through Eq.~(\ref{eq:r-s-QF}) directly.
In view of the equivalence for $f_1(1285)-f_1(1420)$ mixing in the QF basis and the SO basis,
the above ratio defined in Eq.~(\ref{eq:r-s-QF}) can also be expressed in the SO basis
as follows,
\beq
{\rm R_s^{SO}} &\equiv&   \frac{Br(B_s \to J/\psi f_1(1285))}
  {Br(B_s \to J/\psi f_1(1420))} \non
  &=& \frac{\Phi_{f_1(1285)}}{\Phi_{f_1(1420)}} \cdot
\frac{ \left | \frac{\cos\theta_{^3P_1}}{\sqrt{3}}
  \cdot {\cal A}(B_s \to J/\psi f_1) -2 \cdot \frac{\sin\theta_{^3P_1}}{\sqrt{6}}
  \cdot {\cal A}(B_s \to J/\psi f_8) \right|^2}
  { \left |-\frac{\sin\theta_{^3P_1}}{\sqrt{3}}
  \cdot {\cal A}(B_s \to J/\psi f_1)
  -2 \cdot \frac{\cos\theta_{^3P_1}}{\sqrt{6}}
  \cdot {\cal A}(B_s \to J/\psi f_8) \right |^2}\;,
  \label{eq:r-s-SO}
 \eeq
which can also be used to extract out the mixing angle $\theta_{^3P_1}$ approximately
based on the assumption~\cite{Carvalho:2002fh} that ${\cal A}(B_s \to J/\psi f_1) \approx {\cal A}(B_s \to J/\psi f_8)$~
\footnote{ Actually, as presented in Ref.~\cite{Yang:2007zt}, the two SO states $f_1$ and $f_8$
have the similar hadronic parameters, which can also
be seen from the similarity of the relevant input parameters in Eq.~(\ref{eq:mass})
and the related phenomenological discussions.},
then extract out the mixing angle $\phi_{^3P_1}$ via the relation $\phi_{^3P_1}= \theta_i - \theta_{^3P_1}$.

And what's more, 
the decays of $B$ mesons into final states containing the $J/\psi$ charmonium
state also play a special role in studies of CP violation physics~\cite{Abe:2001oa}.
As discussed in the literature~\cite{Yang:2007zt,Cheng:2007mx,Yang:2008xw,Cheng:2008gxa},
the behavior of $1^3P_1$ axial-vector meson is similar with that of the vector meson.
It is naturally expected that the $B_s \to J/\psi f_1(1285)$ and $B_s \to J/\psi f_1(1420)$ decays can serve as the
alternative channels to reduce the errors in the determination of the $B_s-\bar B_s$ mixing
phase $\phi_s$ effectively.

We here will investigate the $B_s \to J/\psi f_1(1285)$ and $B_s \to J/\psi f_1(1420)$ decays in
the perturbative QCD(pQCD) 
approach~\cite{Keum:2000ph,Keum:2000wi,Lu:2000em}
with the aforementioned two mixing schemes.
Because of the similar behavior between the $1^3P_1$ axial-vector mesons
and the vector mesons and the global agreement between the theoretical
predictions in the pQCD approach and the presently existing experimental
data for the $B \to J/\psi V$ decays~\cite{Liu:2013nea},
we can therefore calculate the decay amplitudes for the $B_s \to J/\psi f_1$
and $J/\psi f_8$ decays or $B_s \to J/\psi f_{1s}$ decay at next-to-leading order of
the strong coupling constant $\alpha_s$ straightforwardly
by substituting the kinematic variables and distribution amplitudes of $\phi$ in
the $B_s \to J/\psi \phi$ mode to those of $f_1$ and
$f_8$ or $f_{1s}$ in the considered decays, apart from an overall minus sign that arising from the
definitions of the wave functions for axial-vector and vector mesons.

Since the vector meson $\rho$ and $\omega$ have the same distribution
amplitudes, except for the different decay constant $f_\rho$ and $f_\omega$,
we assume that the distribution amplitude of the QF state $f_{1q}$
is the same one as $a_1(1260)$ with decay constant 
$f_{f_{1q}}= 0.193^{+0.043}_{-0.038}$~GeV
~\cite{Verma:2011yw}.
For the $f_{1s}$ state, for the sake of simplicity,
we adopt the same distribution amplitude as $f_1$
with decay constant $f_{f_{1s}}= 0.230 \pm 0.009$~GeV~\cite{Verma:2011yw}.
In fact, we have confirmed that the CP-averaged branching ratios just vary
$3\%$ for the change of the distribution amplitude of $f_1$ into that of $f_8$.

The following input parameters, such as the QCD scale~({\rm GeV}),
masses~({\rm GeV}), decay constants (GeV) and $B_s$ meson lifetime ({\rm ps})
as given in Refs.~\cite{Beringer:1900zz,Yang:2007zt,Yang:2010ah,Verma:2011yw},
will be used in the numerical calculations:
\beq
 \Lambda_{\overline{\mathrm{MS}}}^{(f=4)} &=& 0.287\;,
 \qquad m_W = 80.41\;,  \qquad m_{b} = 4.8\;,
 \qquad m_{B_s} = 5.37\;;\non
   m_{J/\psi} &=& 3.097\;,\hspace{0.08cm} \qquad m_{f_1} = 1.28 \;,
   \hspace{0.06cm} \qquad   m_{f_8}= 1.29\;,
  \qquad m_{c} = 1.50\;; \non
   f_{J/\psi}&=& 0.405\;,\hspace{0.17cm} \qquad f_{B_s} = 0.23 \;,
   \qquad \hspace{0.17cm} f_{f_1} = 0.245\;,\hspace{-0.15cm} \qquad f_{f_8} = 0.239 \;;\non
\tau_{B_s}&=& 1.497\;
   \hspace{0.25cm} \qquad f_{f_{1s}} = 0.230\;, \qquad  \theta_{^3P_1} = 20^\circ\;, \qquad
   \phi_{^3P_1} = 15.3^\circ.
 \label{eq:mass}
\eeq
For the mixing angle of $f_1(1285)-f_1(1420)$ system, we here adopt
recently updated value $\theta_{^3P_1} \approx 20^\circ$ and
$\phi_{^3P_1} \approx 15.3^\circ$ extracted from the $f_1(1285) \to
\rho \gamma, \phi \gamma$ decays ~\cite{Yang:2010ah},  to calculate the
physical quantities for the two considered $B_s$ decays.
For the Cabibbo-Kobayashi-Maskawa (CKM) matrix elements, we adopt the
Wolfenstein parametrization up to corrections of ${\cal O}(\lambda^5)$
and the updated parameters $A=0.811$,  $\lambda=0.22535$,
$\bar{\rho}=0.131^{+0.026}_{-0.013}$ and $\bar{\eta}=0.345^{+0.013}_{-0.014}$
as given in PDG 2012~\cite{Beringer:1900zz}.

The pQCD predictions for the CP-averaged branching ratios of the
$B_s \to J/\psi f_1(1285)$ and $B_s \to J/\psi f_1(1420)$ decays
within errors in the standard model with the two mixing schemes
are the following:
\begin{itemize}
\item In the QF basis 
\beq
Br(B_s \to J/\psi f_1(1285))  &=& 7.70^{+2.30}_{-1.74}(\omega_{B})
^{+1.05}_{-0.99}(f_M)
^{+3.33}_{-2.50}(a_{i})
^{+1.22}_{-1.25}(m_c)
^{+4.38}_{-3.45}(\phi_{^3P_1})
^{+0.22}_{-0.30}(a_t)\non
&=&\Bigl[7.70^{+6.18}_{-4.88}\Bigr] \times  10^{-5}  \label{eq:brpsif1285-qf} \;,\\
Br(B_s \to J/\psi f_1(1420)) &=&
0.97^{+0.30}_{-0.21}(\omega_{B})
^{+0.14}_{-0.12}(f_M)
^{+0.42}_{-0.31}(a_{i})
^{+0.17}_{-0.15}(m_c)
^{+0.04}_{-0.04}(\phi_{^3P_1})
^{+0.04}_{-0.04}(a_t)\non
&=&\Bigl[0.97^{+0.56}_{-0.42}\Bigr]
\times  10^{-3}   \label{eq:brpsif1420-qf}\;;
\eeq

\item In the SO basis 
\beq
Br(B_s \to J/\psi f_1(1285))  &=&
8.71^{+2.59}_{-1.99}(\omega_{B})
^{+2.46}_{-2.23}(f_M)
^{+9.26}_{-5.40}(a_{i})
^{+1.25}_{-1.34}(m_c)
^{+4.96}_{-3.91}(\theta_{^3P_1})
^{+0.23}_{-0.34}(a_t)\non
&=&\Bigl[8.71^{+11.17}_{-7.44}\Bigr]
\times  10^{-5}  \label{eq:brpsif1285-so} \;,\\
Br(B_s \to J/\psi f_1(1420)) &=&
1.06^{+0.32}_{-0.23}(\omega_{B})
^{+0.16}_{-0.14}(f_M)
^{+0.31}_{-0.25}(a_{i})
^{+0.19}_{-0.18}(m_c)
^{+0.04}_{-0.04}(\theta_{^3P_1})
^{+0.04}_{-0.04}(a_t)\non
&=&\Bigl[1.06^{+0.51}_{-0.41}\Bigr]
\times  10^{-3}   \label{eq:brpsif1420-so}\;;
\eeq
\end{itemize}
where the total errors are obtained by adding the errors from different sources
in quadrature.
The individual theoretical errors are induced by the variation of the
shape parameter $\omega_B = 0.50 \pm 0.05$~GeV~\cite{Ali:2007ff} for the $B_s$ meson
wave function, of the $J/\psi$ meson decay constant $f_{J/\psi}= 0.405
\pm 0.014$~GeV~\cite{Bondar:2004sv,Chen:2005ht} and the $f_1(f_8)$ state decay constant $f_{f_1}= 0.245
\pm 0.013 (f_{f_8}= 0.239 \pm 0.013)$~GeV~\cite{Yang:2007zt} or the $f_{1s}$ state decay constant
$f_{f_{1s}}= 0.230 \pm 0.009$~GeV~\cite{Verma:2011yw}, of the Gegenbauer moments
$a_2^{\parallel}= -0.04 \pm 0.03$ and $a_1^{\perp}= -1.06 \pm 0.36$
($a_2^{\parallel}= -0.07 \pm 0.04$ and $a_1^{\perp}= -1.11 \pm 0.31$)
for the $f_1(f_8)$ distribution amplitudes~\cite{Yang:2007zt}, of the charm quark mass
$m_c =1.50 \pm 0.15$~GeV, and of the mixing angle $\phi_{^3P_1}=(15.3 \pm 4)^\circ$
or $\theta_{^3P_1}= (20 \pm 4)^\circ$ in the QF or SO basis~\cite{Yang:2010ah},
respectively. Moreover, as displayed in Eqs.~(\ref{eq:brpsif1285-qf}-\ref{eq:brpsif1420-so}), the higher order
contributions are also simply investigated by varying the hard scale
$t_{\rm max}$ from $0.8 t$ to $1.2 t$ (not changing $1/b_i, i= 1,2,3$) in the
hard kernel, which has been counted as one of the sources of theoretical
uncertainties.
It is found that the higher order corrections to these considered
$B_s \to J/\psi f_1(1285)$ and $B_s \to J/\psi f_1(1420)$ decays are indeed
small as the naive expectation.
It is worthwhile to stress  that the variation of the CKM parameters has almost no
effects to the CP-averaged branching ratios and polarization fractions of
these considered $B_{s} \to J/\psi f_1(1285)$
and $B_s \to J/\psi f_1(1420)$ decays in the pQCD approach and thus
will be neglected in the numerical results
as shown in Eqs.~(\ref{eq:brpsif1285-qf}-\ref{eq:brpsif1420-so}) and
Table~\ref{tab:pf-bs}.

It is easy to see that the pQCD predictions for the branching ratios
$Br(B_s \to J/\psi f_1(1285))$, in both the SO and QF mixing schemes,
agree well with currently available data $(7.14^{+1.36}_{-1.41}) \times 10^{-5}$
~\cite{Aaij:2013rja} within the theoretical errors.
Meanwhile, we observe that the pQCD predictions
for the branching ratios of $B_s \to J/\psi f_1(1420)$ decay mode are
at the order of $10^{-3}$, very similar to the decay rate of
$B_s \to J/\psi \phi$,  and can be accessed and tested easily
at the running LHCb and forthcoming Super-B experiments in the near future.
The slightly larger central value of $Br(B_s \to J/\psi f_1(1285))$
and $Br(B_s \to J/\psi f_1(1420))$ 
in the SO basis
than the one in the QF basis  is due to the larger decay constants of
$f_1$ and $f_8$ than that of $f_{1s}$, which can be clearly seen in Eq.~(\ref{eq:mass}).

When the very recently measured value of the mixing angle
$\phi_{^3P_1} = 24^\circ$~\cite{Aaij:2013rja} is used in the numerical
calculations, we find the pQCD predictions for the branching ratios:
$Br(B_s \to J/\psi f_1(1285))=18.29 \times 10^{-5}$ ($20.71 \times 10^{-5}$)
and $ Br(B_s \to J/\psi f_1(1420))=0.87 \times 10^{-3}$  ($0.95 \times 10^{-3}$)
in the QF (SO) basis.
One can see that the central values of the above pQCD predictions for the
decay rates $Br(B_s \to J/\psi f_1(1285))$ in both mixing schemes
exceed the measured value as listed in Eq.~(\ref{eq:psif12-ex}).

Moreover, according to the theoretical predictions in the pQCD approach,
one can see that the decay rate for $B_s \to J/\psi f_1(1285)$ is more sensitive to the
variation of the mixing angle $\theta_{^3P_1}(\phi_{^3P_1})$ than that
for $B_s \to J/\psi f_1(1420)$, since the $f_1(1285)$ meson is dominated by the $u\bar u
+ d\bar d$ component while the $f_1(1420)$ meson is determined by the
$s\bar s$ component.

With the help of Eq.~(\ref{eq:r-s-QF}), by combining the decay rate
of $B_s \to J/\psi f_1(1285)$ as given in Eq.~(\ref{eq:psif12-ex})
and $\tan^2\phi = 0.1970 \pm 0.053^{+0.014}_{-0.012}$~\cite{Aaij:2013rja},
one can find that
$Br(B_s \to J/\psi f_1(1420))=(3.42^{+1.15}_{-1.16}) \times 10^{-4}$, which is only
about $35\%$ of our pQCD predictions in both mixing schemes as given in Eqs.~(\ref{eq:brpsif1420-qf}) and (\ref{eq:brpsif1420-so}).
Once the future measurements confirm this estimation, it may imply
the existence of large exotic gluonic component in the $f_1(1420)$ meson, something
similar with the case of $\eta'$ ~\cite{Liu:2012ib} in the $\eta-\eta'$ mixing system,
which would need further studies in the future, although
there are now no any signals observed at the experiments.

Based on the above theoretical predictions for the CP-averaged branching
ratios of $B_s \to J/\psi f_1(1285)$ and $B_s \to J/\psi f_1(1420)$ decays
in the pQCD approach, the ratios of the
decay rates between these two modes can be obtained directly as follows
\beq
{\rm R_s^{QF;th.}} &\equiv&
\frac{Br(B_s \to J/\psi f_1(1285))}{Br(B_s \to J/\psi f_1(1420))}
= 0.079^{+0.078}_{-0.061} \;,
\label{eq:r-s-QF-th}
\eeq
and
\beq
{\rm R_s^{SO;th.}} &\equiv&
\frac{Br(B_s \to J/\psi f_1(1285))}{Br(B_s \to J/\psi f_1(1420))}
= 0.082^{+0.113}_{-0.077} \;,
\label{eq:r-s-SO-th}
\eeq
where  we have kept the masses of $f_1(1285)$ and $f_1(1420)$ mesons in the
phase space factors for the $B_s \to J/\psi f_1(1285)$
and $B_s \to J/\psi f_1(1420)$ decay rates.
The good consistency between these two ratios ${\rm R_s^{QF;th.}}$ and
${\rm R_s^{SO;th.}}$ verify the equivalence of the QF basis and SO basis
for the $f_1(1285)-f_1(1420)$ mixing in the pQCD calculations.
Therefore,
one can extract out the mixing angle $\phi_{^3P_1}$ from the ratio of the
branching ratios for $B_s \to J/\psi
f_1(1285)$ and $B_s \to J/\psi f_1(1420)$ modes in the SO
basis theoretically. The mixing angles for the $f_1(1285)-f_1(1420)$
system extracted through Eq.~(\ref{eq:r-s-QF}) are
$\phi_{^3P_1}=(15.3^{+13.8}_{-12.1})^\circ$ in the QF basis and
$\phi_{^3P_1}=(15.5^{+17.3}_{-14.2})^\circ$ in the SO basis, respectively.
Here, we should point out that the errors induced by the variation of
the input mixing angle are not considered in the extraction of the QF mixing
angle $\phi_{^3P_1}$.
The tiny deviation between the central values of these two QF mixing angles
arises from the very small differences between the decay amplitudes ${\cal A}(B_s \to J/\psi f_1)$
and ${\cal A}(B_s \to J/\psi f_8)$ in the SO basis.

Moreover, within the still large theoretical uncertainties from the
non-perturbative inputs in the pQCD approach, our extracted mixing angle
$\phi_{^3P_1}$ is basically in agreement
with the earlier determination $(15^{+5}_{-10})^\circ$ by Mark-II detector at
SLAC~\cite{Gidal:1987bn}, the updated Lattice QCD analysis
$(21 \pm 5)^\circ$~\cite{Dudek:2013yja}, as well as the preliminary
$(24.0^{+3.2}_{-2.7})^\circ$ reported by the LHCb  Collaboration~\cite{Aaij:2013rja}.
Strictly speaking, the non-perturbative inputs for the involved hadrons need
stringent constraints from the experimental measurements,
which then makes the relevant predictions theoretically reliable
and comparable to the data. Of course, we know that the
precision determination of the mixing angle in $f_1(1285)-f_1(1420)$ system
demands enough data samples collected from various processes.

We have also computed the CP-averaged polarization fractions for $B_s \to J/\psi
f_1(1285)$ and $B_s \to J/\psi f_1(1420)$ decay modes in the pQCD approach.
The numerical results for the polarization fractions are presented
in Table~\ref{tab:pf-bs}, in which various errors induced by the input parameters
have been added in quadrature.

\begin{table*}[htb]
\caption{ The theoretical predictions for the CP-averaged polarization fractions
of $B_s \to J/\psi f_1(1285)$
and $J/\psi f_1(1420)$ decays in the pQCD approach with different mixing schemes.}
\label{tab:pf-bs}
\begin{center}\vspace{-0.2cm}
{ \begin{tabular}[t]{l|l|l|c}
 \hline \hline
 Decay modes     & QF basis(\%) \qquad\qquad & SO basis (\%) \qquad\qquad &  data    \\
\hline
$B_s \to J/\psi f_1(1285)$
&
$\begin{array}{ll}
34.3^{+14.7}_{-9.9}(L) \\
40.7^{+6.2}_{-8.7}(\parallel)\\
24.9^{+3.8}_{-5.8}(\perp)
\end{array} $
&
$\begin{array}{ll}
36.3^{+37.2}_{-17.7}(L)\\
39.8^{+10.4}_{-22.5}(\parallel)\\
23.9^{+7.3}_{-15.5}(\perp)
\end{array} $
&
-
\\
\hline \hline
$B_s \to J/\psi f_1(1420)$
&
$\begin{array}{ll}
34.7^{+14.3}_{-10.0}(L) \\
42.5^{+6.7}_{-9.0}(\parallel)\\
22.8^{+3.5}_{-5.3}(\perp)
\end{array} $
&
$\begin{array}{ll}
33.9^{+9.8}_{-8.5}(L) \\
42.7^{+5.9}_{-6.3}(\parallel)\\
23.4^{+2.7}_{-3.5}(\perp)
\end{array} $
&
-
\\
\hline \hline
\end{tabular}}
\end{center}
\end{table*}

From the pQCD predictions as listed in Table~\ref{tab:pf-bs}, one can see the high
similarity between the theoretical predictions for the three kinds of
polarizations obtained for these two decay modes, and also for the two different
mixing schemes.
Another point is that, in the pQCD approach,
the transverse polarization contributions  dominate these two decays in the QF basis
and the longitudinal polarization fractions are $(24.4 \sim 49.0)\%$ for $B_s \to J/\psi
f_1(1285)$ decay and $(24.7 \sim 49.0)\%$
for $B_s \to J/\psi f_1(1420)$ decay(See Table~\ref{tab:pf-bs}), respectively,
which seems slightly different from that for $B_s \to J/\psi \phi$
channel~\cite{Liu:2013nea}.
Meanwhile, as can be seen from Table~\ref{tab:pf-bs}, the polarization
fractions calculated
in the SO basis indicate that $B_s \to J/\psi f_1(1285)$ decay possibly
has a little large longitudinal contributions when the large theoretical
errors induced by the less constrained
hadronic parameters are taken into account. The above theoretical predictions for the
CP-averaged polarization fractions and the related phenomenology in both mixing schemes
can be tested by the near future experiments at LHCb and/or Super-B.


In summary, motivated by the very recent LHCb measurement on the $B_s \to J/\psi
f_1(1285)$ decay and encouraged by the good agreement between the
pQCD predictions and the available data for the $B \to J/\psi V$ decays,
we studied the $B_s \to J/\psi f_1(1285)$ and $B_s \to J/\psi f_1(1420)$
decays for the first time within the framework of the pQCD 
approach
by including higher order QCD corrections.
We made the first pQCD evaluation for the CP-averaged
branching ratios for the considered $B_s \to J/\psi
f_1(1285)$ and $B_s \to J/\psi f_1(1420)$ decays.
The results arising from a smaller angle $\phi_{^3P_1} \approx 15^\circ$ turn out
to be well consistent with the current measurements within theoretical errors.
By employing the ratio of
the decay rates for the considered two modes, we extracted out the
mixing angle $\phi_{^3P_1}$ of $f_1(1285)-f_1(1420)$ system as
$\phi_{^3P_1}=(15.3^{+13.8}_{-12.1})^\circ$
and $(15.5^{+17.3}_{-14.2})^\circ$ in the QF and SO mixing basis, which are
basically consistent with currently available measurements or
estimations within still large theoretical errors.
Furthermore, the large transverse polarization fractions for these two decay modes
are also predicted for tests by the LHCb and the forthcoming Super-B
experiments. Finally, it is noted that the pQCD predictions
for the considered decays still suffer from large theoretical errors
induced by the uncertainties of the input parameters such as
hadron decay constants and Gegenbauer moments in the distribution
amplitudes of axial-vector states, which are expected to be
constrained by more precision data from various channels in the future.


\begin{acknowledgments}
Xin Liu thanks Sheldon Stone for his useful comments.
This work is supported by the National Natural Science
Foundation of China under Grants Nos.~11205072 and~11235005,
and by a project funded by the Priority Academic Program Development
of Jiangsu Higher Education Institutions (PAPD),
and by the Research Fund of Jiangsu Normal University under Grant No.~11XLR38.
\end{acknowledgments}



\end{document}